\documentclass[twocolumn,conference]{IEEEtran}
\usepackage[T1]{fontenc}
\usepackage[latin9]{inputenc}
\usepackage{color}
\usepackage{prettyref}
\usepackage{amsmath}
\usepackage{amssymb}
\usepackage{graphicx}
\usepackage[unicode=true,
 bookmarks=false,
 breaklinks=false,pdfborder={0 0 0},pdfborderstyle={},backref=false,colorlinks=false]
 {hyperref}
\hypersetup{pdftitle={Your Title},
 pdfauthor={Your Name},
 pdfpagelayout=OneColumn, pdfnewwindow=true, pdfstartview=XYZ, plainpages=false}

\makeatletter
\usepackage{amsmath}
\usepackage{liulyx}
\usepackage{acronym}
\AtBeginDocument{\acrodef{ASP}{antenna separation product}
\acrodef{AWGN}{additive white Gaussian noise}
\acrodef{BEP}{bit error probability}
\acrodef{BER}{bit error rate}
\acrodef{BF-MIMO}[BF\mbox{-}MIMO]{beamforming MIMO}
\acrodef{BF}{beamforming}
\acrodef{bpcu}{bits per channel use}
\acrodef{CP}{cyclic prefix}
\acrodef{CSI}{channel state information}
\acrodef{CSIR}{channel state information at RX}
\acrodef{SSK}{space shift keying}
\acrodef{CSIT}{channel state information at TX}
\acrodef{DCMC}{discrete\mbox{-}input continuous\mbox{-}output memoryless channel}
\acrodef{DFT}{discrete Fourier transform}
\acrodef{DL-TR-GSM}{dual-layered transmit-receive \acl{GSM}}
\acrodef{DLT}{dual-layered transmission}
\acrodef{EGC}{equal gain combining}
\acrodef{FSPL}{free space path loss}
\acrodef{FFT}{fast Fourier transform}
\acrodef{FDE}{frequency domain equalization}
\acrodef{GRSM}{generalized \acl{RSM}}
\acrodef{GSM}{generalized \acl{SM}}
\acrodef{IFFT}{invserse fast Fourier transform}
\acrodef{ICI}{inter-channel interference}
\acrodef{iid}[i.i.d.]{independent and identically distributed}
\acrodef{IQ}{in\mbox{-}phase and quadrature}
\acrodef{ISI}{intersymbol interference}
\acrodef{ISI-free}[ISI\mbox{-}free]{intersymbol interference free}
\acrodef{LOS}{line\mbox{-}of\mbox{-}sight}
\acrodef{mmWave}{millimeter-wave}
\acrodef{MIMO}{multiple\mbox{-}input multiple\mbox{-}output}
\acrodef{MISO}{multiple input single output}
\acrodef{ML}{maximum likelihood}
\acrodef{MRC}{maximal ratio combining}
\acrodef{MMSE}{minimum mean square error}
\acrodef{MU-TR-GSM}{multiuser transmit-receive  \acl{GSM} }
\acrodef{NCSIT}{no channel state information at TX}
\acrodef{NLOS}{non\mbox{-}\acs{LOS}} 
\acrodef{OFDM}{orthogonal frequency division multiplexing}
\acrodef{PA}{power amplifier}
\acrodef{PAE}{power added efficiency}
\acrodef{PAPR}{peak\mbox{-}to\mbox{-}average power ratio}
\acrodef{PDF}{probability density function}
\acrodef{PEP}{pairwise error probability}
\acrodefplural{PEP}{pairwise error probabilities}
\acrodef{PMP}{probability mass function}
\acrodef{PSM}{precoding-aided spatial modulation}
\acrodef{QSM}{quadrature spatial modulation}
\acrodef{RC}{reorganization computation}
\acrodef{RSM}{receive spatial modulation}
\acrodef{RX}{receiver}
\acrodef{SEP}{symbol error probability}
\acrodef{SER}{symbol error rate}
\acrodef{SM}{spatial modulation}
\acrodef{SMX-MIMO}[SMX\mbox{-}MIMO]{spatial multiplexing MIMO}
\acrodef{SMX}{spatial multiplexing}
\acrodef{SNR}{signal-to-noise ratio}
\acrodef{SC}{single carrier}
\acrodef{SVD}{singular value decomposition}
\acrodef{SPST}{single pole single-throw}
\acrodef{TDE}{time domain equalization}
\acrodef{TX}{transmitter}
\acrodef{ULA}{uniform linear array}
\acrodef{ZF}{zero-forcing}
\acrodef{ZMCG}{zero-mean complex Gaussian}

}
 
\PassOptionsToPackage{bookmarks=false}{hyperref}

\makeatother

\begin{document}
\title{Transmit-Receive Generalized Spatial Modulation Based on Dual-layered
MIMO Transmission}
\author{\IEEEauthorblockN{Nemanja~Stefan~Perovi\'c\IEEEauthorrefmark{1}, Marco~Di~Renzo\IEEEauthorrefmark{2},
and Mark~F.~Flanagan\IEEEauthorrefmark{1}}\IEEEauthorblockA{\IEEEauthorrefmark{1}School of Electrical and Electronic Engineering,
University College Dublin\\
Belfield, Dublin 4, Ireland\\
Email: nemanja.stefan.perovic@ucd.ie and mark.flanagan@ieee.org}\IEEEauthorblockA{\IEEEauthorrefmark{2}Laboratoire des Signaux et Syst\`emes, CNRS,
CentraleSup\'elec, Univ Paris Sud, Universit\'e Paris-Saclay\\
3 rue Joliot Curie, Plateau du Moulon, 91192 Gif-sur-Yvette, France\\
Email: marco.direnzo@l2s.centralesupelec.fr}}
\maketitle
\begin{abstract}
We propose a novel scheme for downlink multiuser \ac{MIMO} systems,
called \ac{DL-TR-GSM}. The proposed scheme is based on the concept
of \ac{DLT} which uses two receive antenna power levels instead of
receive antenna activation/inactivation to transmit data in the receive
spatial domain. Hence, in order to minimize the \ac{BER} for \ac{DL-TR-GSM},
the optimal ratio between the two power levels is determined. To further
characterize \ac{DL-TR-GSM}, we fully derive the computational complexity
and show a significant computational complexity reduction as well
as a required hardware complexity reduction of \ac{DL-TR-GSM}, compared
to a state-of-the-art benchmark scheme. Simulation results confirm
the performance advantages of \ac{DL-TR-GSM}.\textcolor{black}{\acresetall{}}
\end{abstract}

\begin{IEEEkeywords}
Dual-layered transmission, \ac{GSM}, \ac{MIMO}, multiuser communications.
\textcolor{black}{\acresetall{}}
\end{IEEEkeywords}

\section{Introduction}

Among many existing \ac{MIMO} schemes, \ac{SM} has attracted a lot
of research interest in recent years. In \ac{SM}, two data streams
are transmitted \textemdash{} one in the conventional \ac{IQ} domain
(by employing e.g. PSK or QAM modulation), and the other in the so-called
spatial domain by selecting and activating one from all available
transmit antenna \cite{Mesleh2008,DiRenzo2014}. A straightforward
extension of SM is to allow activation of more than one transmit antennas
per time slot and possibly also to transmit more than one \ac{IQ}
stream simultaneously \cite{Younis2010}. The extended scheme is called
\ac{GSM}.

Recently, a scheme which is operationally dual to SM was developed,
called \ac{RSM} \cite{Yang2011,Perovic2015a}. The main difference
between SM and RSM comes from the signal transmission in the spatial
domain, where RSM transmits data by selecting one out of all available
\emph{receive} antennas. Accordingly, this antenna is used for the
reception of the transmitted \ac{IQ} stream.  Similarly, the concept
of \ac{RSM} may be extended by selecting more than one receive antenna
per time slot for the reception of multiple \ac{IQ} stream transmission
\cite{Zhang2013,Zhang2015}. This scheme is called \ac{GRSM}.

Another interesting extension of \ac{RSM} is that of \ac{DLT} \cite{Masouros2016}.
In contrast to conventional \ac{RSM}/\ac{GRSM} which utilizes a
subset of the receive antennas, \ac{DLT} uses all available receive
antennas for the reception of the transmitted \ac{IQ} streams. Consequently,
\ac{DLT} requires a new approach to transmit information in the spatial
domain and thus  DLT applies two power levels to distinguish the
\textquotedblleft selected\textquotedblright{} from the ``non-selected''
receive antennas in the spatial domain \cite{Masouros2016}. In this
way, the spatial symbols are encoded onto the signal power levels
at the receive antennas.

Although the basic theory for \ac{SM} and \ac{RSM} was initially
developed for single-user communication systems, an increasing number
of research works consider their application in multiuser scenarios.
In \cite{Narasimhan2015}, the authors considered a multiuser uplink
transmission scheme with SM implemented at each user. To enable SM
in multiuser downlink communications, a closed-form precoding solution
was derived in \cite{Narayanan2014}. In \cite{Humadi2014}, an implementation
of \ac{RSM}/\ac{GRSM} in massive \ac{MIMO} systems was investigated.
A more detailed analysis of \ac{GRSM} for multiuser downlink communications
was presented in \cite{Stavridis2016}.

The papers listed above consider multiuser communication schemes that
are based on the \ac{SM} or the \ac{RSM} operation principle. To
the best of the authors' knowledge, the only multiuser scheme that
simultaneously supports the \ac{SM} and \ac{RSM} operation principle
is presented in \cite{Pizzio2016}, and is called \ac{MU-TR-GSM}.
In each time slot, the base station in \ac{MU-TR-GSM} selects a subset
of the transmit antennas to be active. From those antennas, the base
station transmits IQ streams to the users. Also, this antenna activation
enables \ac{MU-TR-GSM} to send data in the transmit spatial domain.
Each user receives the transmitted IQ streams by a subset of receive
antennas, whose selection enables \ac{MU-TR-GSM} to send data in
the receive spatial domain. Therefore, \ac{MU-TR-GSM} manages to
combine the principles of operation of \ac{SM} and \ac{RSM}. Despite
the advantage of combining the SM and the \ac{RSM} operation principles,
\ac{MU-TR-GSM} requires a high computational complexity (see Section
\prettyref{subsec:Com-Compl}) which presents a significant barrier
to its practical implementation.

Against this background, the contributions of this paper are listed
as follows:
\begin{enumerate}
\item We propose a novel multiuser communication scheme, called \ac{DL-TR-GSM},
that simultaneously supports the operation of the \ac{SM} and \ac{RSM}
operation principles. However, in contrast to \ac{MU-TR-GSM} which
applies the conventional \ac{RSM} transmission, \ac{DL-TR-GSM} is
based on \ac{DLT}.
\item We show, through a detailed computational complexity analysis and
through simulations, that \ac{DL-TR-GSM} enables a considerable computational
complexity reduction, at the cost of a minor degradation in the \ac{BER}
performance.
\item We also provide a hardware complexity analysis which demonstrates
that \ac{DL-TR-GSM} requires a lower number of RF chains at the receiver
compared to \ac{MU-TR-GSM}. As a result, \ac{DL-TR-GSM} provides
a large reduction of the receive power consumption at each user.
\item We introduce a low-complexity detector for \ac{DL-TR-GSM}, referred
to as the separate detector. Simulation results show that this detector
provides a very similar \ac{BER} to that provided by the optimal
\ac{ML} detector.
\end{enumerate}

\section{System Model}

\subsection{\ac{DL-TR-GSM}}

\begin{figure}[t]
\begin{centering}
\includegraphics{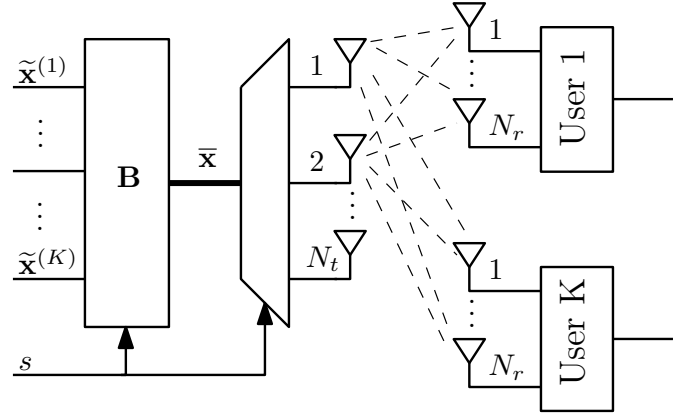}
\par\end{centering}
\caption{Block diagram for the proposed scheme. \label{fig:Diagram-Multi}}
\end{figure}

The block diagram for the considered \ac{DL-TR-GSM} scheme is shown
in \prettyref{fig:Diagram-Multi}. It depicts a downlink communication
scenario between a base station equipped with $N_{t}$ transmit antennas
and $K$ users equipped with $N_{r}$ receive antennas per user. Accordingly,
the channel matrix of the \ac{DL-TR-GSM} scheme can be expressed
as
\[
\mathbf{H}=\left[\mathbf{H}^{(1)^{\mathrm{T}}}\;\mathbf{H}^{(2)^{\mathrm{T}}}\;\cdots\;\mathbf{H}^{(K)^{\mathrm{T}}}\right]^{\mathrm{T}},
\]
where $\mathbf{\mathbf{H}}^{(k)}\in\mathbb{C}^{N_{r}\times N_{t}}$
is the channel matrix between the base station and the $k$-th user.

In each time slot, the base station activates a subset of $N_{\mathrm{tact}}$
($KN_{r}\le N_{\mathrm{tact}}<N_{t}$) transmit antennas, which form
1 out of $N_{\mathrm{tcomb}}=2^{\left\lfloor \log_{2}(N_{t}/N_{\mathrm{tact}})\right\rfloor }$
transmit antenna combinations. For ease of comparison of our later
results with those in \cite{Pizzio2016}, we assume that each transmit
antenna can belong to only one transmit antenna combination; thus,
the data rate in the transmit spatial domain is $\left\lfloor \log_{2}\frac{N_{t}}{N_{\mathrm{tact}}}\right\rfloor $
instead of $\left\lfloor \log_{2}{N_{t} \choose N_{\mathrm{tact}}}\right\rfloor $.
For the \emph{s}-th combination of activated transmit antennas ($s=1,\dots,N_{\mathrm{tcomb}}$),
the resulting channel matrix $\mathbf{H}_{s}$ consists of the $N_{\mathrm{tact}}$
columns of $\mathbf{H}$ that correspond to the active transmit antennas.
Implementing \ac{SVD} on any constituent matrix $\mathbf{H}_{s}^{(k)}$
of $\mathbf{H}_{s}$, we obtain
\begin{equation}
\mathbf{H}_{s}^{(k)}=\mathbf{U}_{s}^{(k)}[\boldsymbol{\Lambda}_{s}^{(k)}\;\mathbf{0}]\left[\begin{array}{c}
\mathbf{V}_{1,s}^{(k)^{\mathrm{H}}}\\
\mathbf{V}_{2,s}^{(k)^{\mathrm{H}}}
\end{array}\right]=\mathbf{U}_{s}^{(k)}\boldsymbol{\Lambda}_{s}^{(k)}\mathbf{V}_{1,s}^{(k)^{\mathrm{H}}},\label{eq:SVD}
\end{equation}
where $\mathbf{U}_{s}^{(k)}\in\mathbb{C}^{N_{r}\times N_{r}}$ is
a unitary matrix, $\boldsymbol{\Lambda}_{s}^{(k)}\in\mathbb{C}^{N_{r}\times N_{r}}$
is a diagonal matrix of singular values and $\mathbf{V}_{1,s}^{(k)}\in\mathbb{C}^{N_{\mathrm{tact}}\times N_{r}}$.
Now, the overall receive signal vector of all $K$ users, $\overline{\mathbf{y}}=[\begin{array}{ccc}
\mathbf{\overline{\mathbf{y}}}^{(1)^{\mathrm{T}}}\;\mathbf{\overline{\mathbf{y}}}^{(2)^{\mathrm{T}}}\; & \cdots & \;\mathbf{\overline{\mathbf{y}}}^{(K)^{\mathrm{T}}}\end{array}]^{\mathrm{T}}\in\mathbb{C}^{KN_{r}\times1}$, can be written as \cite{Liu2009a}
\begin{equation}
\overline{\mathbf{y}}=\mathbf{H}_{s}\overline{\mathbf{x}}+\overline{\mathbf{n}}=\overline{\mathbf{U}}_{s}\overline{\mathbf{\Lambda}}_{s}\overline{\mathbf{V}}_{1,s}^{\mathrm{H}}\overline{\mathbf{x}}+\overline{\mathbf{n}},
\end{equation}
where, according to \prettyref{eq:SVD}, we introduced the definitions
\[
\begin{array}{c}
\overline{\mathbf{U}}_{s}=\mathrm{diag}(\mathbf{U}_{s}^{(1)}\;\mathbf{U}_{s}^{(2)}\;\cdots\;\mathbf{U}_{s}^{(K)})\\
\overline{\mathbf{\Lambda}}_{s}=\mathrm{diag}(\mathbf{\mathbf{\Lambda}}_{s}^{(1)}\;\mathbf{\mathbf{\Lambda}}_{s}^{(2)}\;\cdots\;\mathbf{\mathbf{\Lambda}}_{s}^{(K)})\\
\overline{\mathbf{V}}_{1,s}=[\mathbf{V}_{1,s}^{(1)}\;\mathbf{V}_{1,s}^{(2)}\;\cdots\;\mathbf{V}_{1,s}^{(K)}].
\end{array}
\]
Moreover, $\overline{\mathbf{x}}\in\mathbb{C}^{N_{\mathrm{tact}}\times1}$
is the transmit signal vector of the base station and  $\overline{\mathbf{n}}$
is the noise vector with $KN_{r}$ \ac{iid} elements that are distributed
according to $\mathcal{CN}(0,N_{0})$, where $N_{0}$ denotes the
(one-sided) power spectral density of the additive white Gaussian
noise (AWGN).

To enable downlink signal transmission without inter-channel and inter-user
interference, a precoder is required at the transmitter. Hence, the
transmit \ac{IQ} symbol vector $\tilde{\mathbf{x}}$, which contains
$KN_{r}$ \ac{IQ} symbols of the \emph{M}-PSK modulation alphabet,
is precoded before its transmission, yielding the vector
\[
\overline{\mathbf{x}}=\mathbf{B}\mathbf{\tilde{x}}.
\]
The precoding matrix is defined as

\begin{equation}
\mathbf{B}=\overline{\mathbf{V}}_{1,s}(\overline{\mathbf{V}}_{1,s}^{\mathrm{H}}\overline{\mathbf{V}}_{1,s})^{-1}\overline{\boldsymbol{\beta}}_{s}\overline{\mathbf{P}}
\end{equation}
where the diagonal matrix $\mathbf{\overline{\boldsymbol{\beta}}_{\mathrm{s}}=\mathrm{diag}}(\boldsymbol{\beta}_{s}^{(1)}\;\boldsymbol{\beta}_{s}^{(2)}\;\cdots\;\boldsymbol{\beta}_{s}^{(K)})=\mathrm{diag}(\beta_{s,1}^{(1)}\;\cdots\;\beta_{s,N_{r}}^{(1)}\;\cdots\;\beta_{s,1}^{(K)}\;\cdots\;\beta_{s,N_{r}}^{(K)})$
serves to ensure a constant average transmit power. Assuming that
all diagonal elements in $\overline{\boldsymbol{\beta}}_{\mathrm{s}}$
are equal as in \cite{Liu2009a}, we obtain $\overline{\boldsymbol{\beta}}_{\mathrm{s}}=\beta_{s}\mathbf{I}_{KN_{r}}$,
where
\begin{equation}
\beta_{s}=\sqrt{\frac{KN_{r}}{\mathrm{Tr}\left[(\overline{\mathbf{V}}_{1,s}^{\mathrm{H}}\overline{\mathbf{V}}_{1,s})^{-1}\right]}}.\label{eq:beta}
\end{equation}
In the remainder of the paper, we will assume that this is the case,
and we will refer to $\beta_{s}$ as the \emph{scaling coefficient}.

To transmit data in the receive spatial domain, \ac{DL-TR-GSM} utilizes
the power level matrix $\mathbf{\overline{P}=\mathrm{diag}}(\mathbf{P}^{(1)}\;\mathbf{P}^{(2)}\;\cdots\;\mathbf{P}^{(K)})$.
Each constituent matrix $\mathbf{P}^{(k)}$ ($k\in\{1,\dots,K\}$)
is a diagonal matrix whose $r$-th diagonal element takes the value
$\sqrt{P_{1}}$ if the $r$-th receive antenna of the \emph{k}-th
user is ``non-selected'' or $\sqrt{P_{2}}$ if the $r$-th receive
antenna of the \emph{k}-th user is ``selected''. Indices of the
``selected'' and the ``non-selected'' receive antennas of one
user determine $N_{r}$ bits transmitted to that user in the receive
spatial domain. More precisely, the indices of the ``non-selected''
receive antennas specify the positions of zeros and the indices of
the ``selected'' receive antennas specify the positions of ones.
Hereinafter, we assume $P_{1}<P_{2}$ and the set of all possible
$\mathbf{\mathbf{P}}^{(k)}$ is denoted by $\mathit{\mathcal{P}}$
(hence $\left|\mathit{\mathcal{P}}\right|=2^{N_{r}}$).

From the previous expressions, the receive signal vector of the $k$-th
user can be written as follows:
\begin{align}
\mathbf{y}^{(k)} & =\mathbf{U}_{s}^{(k)}\mathbf{\mathbf{\Lambda}}_{s}^{(k)}\boldsymbol{\beta}_{s}^{(k)}\mathbf{P}_{i}^{(k)}\mathbf{\tilde{x}}_{m}^{(k)}+\mathbf{n}^{(k)}\nonumber \\
 & =\mathbf{G}_{s}^{(k)}\mathbf{P}_{i}^{(k)}\mathbf{\tilde{x}}_{m}^{(k)}+\mathbf{n}^{(k)},
\end{align}
where $m\in\{1,\dots,M^{N_{r}}\}$ and $i\in\{1,\dots,2^{N_{r}}\}$
are the index of the transmitted \ac{IQ} symbol vector and the index
of the used power level matrix for the $k$-th user, respectively.
To cover the data transmission in all three domains we refer to the
column vector $\mathbf{G}_{s}^{(k)}\mathbf{P}_{i}^{(k)}\mathbf{\tilde{x}}_{m}^{(k)}$
as the \emph{supersymbol}, where $\mathbf{G}_{s}^{(k)}=\mathbf{U}_{s}^{(k)}\mathbf{\mathbf{\Lambda}}_{s}^{(k)}\boldsymbol{\beta}_{s}^{(k)}$.
One should note that each supersymbol is uniquely determined by a
particular $(s,i,m)$ index combination. 

Now, the optimal \ac{ML} detector of the \emph{k}-th user is given
as
\begin{align}
\{\hat{s},\hat{i},\hat{m}\} & =\mathop{\mathop{\text{arg min}}}_{\substack{s\in\{1,...,N_{\mathrm{tcomb}}\}\\
m\in\{1,...,M^{N_{r}}\}\\
i\in\mathit{\mathcal{P}}
}
}\left\Vert \mathbf{y}^{(k)}-\mathbf{G}_{s}^{(k)}\mathbf{P}_{i}^{(k)}\mathbf{\tilde{x}}_{m}^{(k)}\right\Vert ^{2},\label{eq:ML_det}
\end{align}
where $\hat{s}$ is the index of the detected transmit antenna combination,
$\hat{i}$ is index of the detected power level matrix and $\hat{m}$
is the index of the detected IQ modulation symbol vector.

Finally, we may note that the data rate per user of \ac{DL-TR-GSM}
is
\begin{equation}
\eta=\left\lfloor \log_{2}\frac{N_{t}}{N_{\mathrm{tact}}}\right\rfloor +N_{r}(1+\log_{2}M).
\end{equation}

\subsection{\ac{MU-TR-GSM}}

As mentioned previously, the main difference between \ac{MU-TR-GSM}
and \ac{DL-TR-GSM} is the data transmission in the receive spatial
domain. In contrast to \ac{DL-TR-GSM} which uses the \ac{DLT},
\ac{MU-TR-GSM} follows the conventional RSM operation principle.
It selects a subset of $N_{\mathrm{ract}}$ $(\text{0}<N_{\mathrm{ract}}<N_{r})$
receive antennas at one user, so that each active receive antenna
in the subset receives one IQ stream. Since there are $N_{\mathrm{rcomb}}=2^{\left\lfloor \log_{2}{N_{r} \choose N_{\mathrm{ract}}}\right\rfloor }$
receive antenna combinations, the data rate in the receive spatial
domain is $\left\lfloor \log_{2}{N_{r} \choose N_{\mathrm{ract}}}\right\rfloor $
bits per user. Another consequence of this change is the construction
of the effective channel matrix $\mathbf{H}_{s}$. Here, $\mathbf{H}_{s}$
is obtained by selecting $N_{\mathrm{tact}}$ columns and $KN_{r}$
rows of $\mathbf{H}$ that correspond to the active transmit and receive
antennas, respectively. Further signal preprocesing at the transmitter
is same as for \ac{DL-TR-GSM}. The only difference is that the precoding
matrix expression does not contain the power level matrix $\overline{\mathbf{P}}$
and that the ratio numerator $KN_{r}$ in \prettyref{eq:beta} should
be replaced by $KN_{\mathrm{ract}}$. At the reception, we execute
the \ac{ML} detection as explained in \cite{Pizzio2016}.

\section{Determining the Optimum Power Levels}

In this section, we derive the \ac{BEP} expression for \ac{DL-TR-GSM}
and based on this we derive the optimal power levels $P_{1}$ and
$P_{2}$. As all the users are assumed to have the same propagation
conditions, the following analytical development is user-independent.
Therefore, the following expressions are valid for an arbitrary user
and we omit the user index in the superscript.

The upper bound for the \ac{BEP} is given by \cite{Liu2018a}
\begin{align}
P_{e} & \le\frac{1}{\eta2^{\eta}}\sum_{(s,i,m)}\sum_{(s_{1},i_{1},m_{1})}D\left((s,i,m)(s_{1},i_{1},m_{1})\right)\nonumber \\
 & \mathbb{\qquad\qquad E_{\mathbf{H}}}\left\{ \mathrm{PEP}\left((s,i,m)(s_{1},i_{1},m_{1})\right)\right\} 
\end{align}
where the indices $(s,i,m)$ and $(s_{1},i_{1},m_{1})$ determine,
respectively, the transmitted and the detected supersymbol. $\mathbb{E_{\mathbf{H}}}\{\mathrm{PEP}((s,i,m)(s_{1},i_{1},m_{1}))\}$
denotes the average \ac{PEP} between the aforementioned mentioned
supersymbols and $D\left((s,i,m)(s_{1},i_{1},m_{1})\right)$ is the
Hamming distance between the binary representations of these supersymbols.
For a given $\mathbf{H}$, if the \ac{ML} detection in \prettyref{eq:ML_det}
is used, the \ac{PEP} can be expressed as
\begin{align*}
 & \mathrm{PEP}\left((s,i,m)(s_{1},i_{1},m_{1})\right)\\
= & \mathrm{Pr}\left[\left\Vert \mathbf{y}-\mathbf{G}_{s}\mathbf{P}_{i}\mathbf{\tilde{x}}_{m}\right\Vert ^{2}>\left\Vert \mathbf{y}-\mathbf{G}_{s_{1}}\mathbf{P}_{i_{1}}\mathbf{\tilde{x}}_{m_{1}}\right\Vert ^{2}\right]\\
= & \mathrm{Pr}\left[\mathfrak{R}\left\{ \mathbf{n}\mathrm{^{H}}\left(\mathbf{G}_{\mathrm{s}_{1}}\mathbf{P}_{i_{1}}\mathbf{\tilde{x}}_{m_{1}}-\mathbf{G}_{s}\mathbf{P}_{i}\mathbf{\tilde{x}}_{m}\right)\right\} \vphantom{>\frac{1}{2}\left\Vert \mathbf{G}_{s_{1}}\mathbf{P}_{\mathrm{i_{1}}}\mathbf{\tilde{x}}_{m_{1}}-\mathbf{G}_{s}\mathbf{P}_{i}\mathbf{\tilde{x}}_{m}\right\Vert ^{2}}\right.\\
 & \phantom{\Pr[}\left.\vphantom{\mathfrak{R}\left\{ \mathbf{n}\mathrm{^{H}}\left(\mathbf{G}_{\mathrm{s}_{1}}\mathbf{P}_{i_{1}}\mathbf{\tilde{x}}_{m_{1}}-\mathbf{G}_{s}\mathbf{P}_{i}\mathbf{\tilde{x}}_{m}\right)\right\} }>\frac{1}{2}\left\Vert \mathbf{G}_{s_{1}}\mathbf{P}_{i_{1}}\mathbf{\tilde{x}}_{m_{1}}-\mathbf{G}_{s}\mathbf{P}_{i}\mathbf{\tilde{x}}_{m}\right\Vert ^{2}\right].
\end{align*}
Since the left-hand side in the previous equation is distributed according
to $\mathcal{N}(0,\left\Vert \mathbf{G}_{s_{1}}\mathbf{P}_{i_{1}}\mathbf{\tilde{x}}_{m_{1}}-\mathbf{G}_{s}\mathbf{P}_{i}\mathbf{\tilde{x}}_{m}\right\Vert ^{2}\cdot N_{0}/2)$,
we get
\begin{multline}
\mathrm{PEP}\left((s,i,m)(s_{1},i_{1},m_{1})\right)=Q\left(\Phi/\sqrt{2N_{0}}\right)=\\
Q\left(\sqrt{\left\Vert \mathbf{G}_{s_{1}}\mathbf{P}_{i_{1}}\mathbf{\tilde{x}}_{m_{1}}-\mathbf{G}_{s}\mathbf{P}_{i}\mathbf{\tilde{x}}_{m}\right\Vert ^{2}}/\sqrt{2N_{0}}\right),\label{eq:PEP}
\end{multline}
where $\Phi$ is the Euclidean distance between the considered supersymbols.

\subsection{Power Ratio $\alpha$}

The ratio of the power levels used for communicating data in the receive
spatial domain is 
\begin{equation}
\alpha=\frac{P_{2}}{P_{1}}\label{eq:cond_1}
\end{equation}
and it satisfies $\alpha>1$. While the chosen power levels need to
maintain the average transmit power unchanged, we have $(P_{1}+P_{2})/2=1.$
Thus we obtain $P_{1}=2/(1+\alpha)$ and $P_{2}=2\alpha/(1+\alpha).$
Since $P_{1}$ and $P_{2}$ are determined entirely by $\alpha$,
the goal is to find the optimal $\alpha$ that ensures the best error
rate performance.

Note that while \prettyref{eq:PEP} captures the PEPs associated with
all possible error events for DL-TR-GSM, for mathematical tractability
reasons we will consider in the following analysis only the individual
PEPs of the IQ domain and of the receive spatial domain. In these
two cases, the Euclidean distance $\Phi$ of \ac{DL-TR-GSM} in \prettyref{eq:PEP}
is mathematically equivalent to the Euclidean distance of a transmit
system that consists of $N_{r}$ parallel orthogonal subchannels.
As the channel gains of these subchannels are proportional to the
singular values in $\mathbf{\mathbf{\Lambda}}_{s}$, the minimum Euclidean
distances, i.e. the maximum \acp{PEP}, will always occur in the subchannel
with the channel gain proportional to the smallest singular value
$\lambda_{s,N_{r}}$.

In the \ac{IQ} domain, due to the use of \emph{M}-PSK modulation,
the maximum \ac{PEP} can be expressed as follows:
\begin{equation}
\mathrm{PEP_{IQ,\max}}=Q\left(\beta_{s}\lambda_{s,N_{r}}\sqrt{\frac{P_{1}}{2N_{0}}}\left|b_{m}-b_{m_{1}}\right|_{\min}\right).\label{eq:PEP_IQ_int}
\end{equation}
For two \emph{M}-PSK symbols $b_{m}$ and $b_{m_{1}}$, we have $\left|b_{m}-b_{m_{1}}\right|_{\min}=2\sin(\pi/M)$
and the previous expression can be re-written as
\begin{equation}
\mathrm{PEP_{IQ,\max}}=Q\left(\beta_{s}\lambda_{s,N_{r}}\sqrt{\frac{2P_{1}}{N_{0}}}\sin\frac{\pi}{M}\right).\label{eq:PEP_IQ}
\end{equation}

From \prettyref{eq:PEP}, the maximum \ac{PEP} of the receive spatial
domain is given as
\begin{equation}
\mathrm{PEP_{RSP,\max}}=Q\left(\beta_{s}\lambda_{s,N_{r}}\frac{\sqrt{P_{2}}-\sqrt{P_{1}}}{\sqrt{2N_{0}}}\right).\label{eq:PEP_SP}
\end{equation}

We define the optimal value of $\alpha$ to be that which minimizes
the maximum PEPs among \prettyref{eq:PEP_IQ} and \prettyref{eq:PEP_SP}.
Therefore, the following equation is valid:
\[
\beta_{s}\lambda_{s,N_{r}}\sqrt{\frac{2P_{1}}{N_{0}}}\sin\frac{\pi}{M}=\beta_{s}\lambda_{s,N_{r}}\frac{\sqrt{P_{2}}-\sqrt{P_{1}}}{\sqrt{2N_{0}}}.
\]
After some simple algebraic manipulations, the optimal power ratio
$\alpha$ is given as
\begin{equation}
\alpha_{\mathrm{opt}}=\left(1+2\sin\frac{\pi}{M}\right)^{2}.\label{eq:alpha_opt}
\end{equation}

\section{System Complexity Analysis}

\subsection{Computational Complexity\label{subsec:Com-Compl}}

In this subsection, we derive the computational complexity of \ac{DL-TR-GSM}
and \ac{MU-TR-GSM}\emph{. }The computational complexity refers to
the number of mathematical operations required for the calculation
of all \acp{SVD} and scaling coefficients $\beta_{s}$ that are needed
in order to perform signal transmission and detection.

In \ac{DL-TR-GSM}, \ac{SVD} is performed for $\mathbf{H}_{s}^{(k)}$
matrices of dimension $N_{r}\times N_{\mathrm{tact}}$, and $4N_{r}^{2}N_{\mathrm{tact}}+22N_{\mathrm{tact}}$
operations are needed for each \ac{SVD} \cite{Golub2012a}. As $N_{\mathrm{tcomb}}$
SVDs are required for each user, the total number of all \acp{SVD}
in \ac{DL-TR-GSM} equals $KN_{\mathrm{tcomb}}$. The complexity
of computing $\beta_{s}$ is primarily determined by the complexity
of the denominator in \prettyref{eq:beta}. Since the matrix $\overline{\mathbf{V}}_{1,s}^{\mathrm{H}}\overline{\mathbf{V}}_{1,s}$
is a $KN_{r}\times KN_{r}$ Hermitian matrix, only the elements on
the main diagonal and below (or above), i.e. $((KN_{r})^{2}+KN_{r})/2$
matrix elements, need to be computed. Since the computation of a single
element requires $2N_{\mathrm{tact}}-1$ operations, the computational
complexity of $\overline{\mathbf{V}}_{1,s}^{\mathrm{H}}\overline{\mathbf{V}}_{1,s}$
is $((KN_{r})^{2}+KN_{r})(N_{\mathrm{tact}}-1/2)$. Inversion of the
aforementioned matrix requires $(KN_{r})^{3}+(KN_{r})^{2}+KN_{r}$
operations \cite{Perovic2015} and the computation of the matrix trace
requires $KN_{r}-1$ operations. In addition, we have 1 square root,
1 multiplication and 1 division operation. The number of different
$\beta_{s}$ values in \ac{DL-TR-GSM} is $N_{\mathrm{tcomb}}$. In
summary, the total computational complexity of \ac{DL-TR-GSM} is
given by
\begin{align*}
C_{\mathrm{DL}} & =KN_{\mathrm{tcomb}}(4N_{r}^{2}N_{\mathrm{tact}}+22N_{\mathrm{tact}})+N_{\mathrm{tcomb}}\left[(KN_{r})^{3}\vphantom{+(KN_{r})^{2}\left(N_{\mathrm{tact}}+\frac{1}{2}\right)+(KN_{r})\left(N_{\mathrm{tact}}+\frac{3}{2}\right)+3)}\right.\\
 & \left.\vphantom{N_{\mathrm{tcomb}}((KN_{r})^{3}}+(KN_{r})^{2}\left(N_{\mathrm{tact}}+\frac{1}{2}\right)+(KN_{r})\left(N_{\mathrm{tact}}+\frac{3}{2}\right)+2\right].
\end{align*}

The computational complexity derivation given above for \ac{DL-TR-GSM}
is applicable with some minor modifications to \ac{MU-TR-GSM}\emph{.
}One difference comes from the fact that the \ac{MU-TR-GSM} scheme
activates $N_{\mathrm{ract}}$ out of $N_{r}$ available receive antennas.
The other difference originates from the number of \acp{SVD} and
$\beta_{s}$ values which are given by $KN_{\mathrm{tcomb}}N_{\mathrm{rcomb}}$
and $N_{\mathrm{tcomb}}N_{\mathrm{rcomb}}^{K}$ respectively for \ac{MU-TR-GSM}.
To summarize, the computational complexity of \ac{MU-TR-GSM} is given
by
\begin{align*}
C_{\mathrm{MU}} & =KN_{\mathrm{tcomb}}N_{\mathrm{rcomb}}\left(4N_{\mathrm{ract}}^{2}N_{\mathrm{tact}}+22N_{\mathrm{tact}}\right)\\
 & +N_{\mathrm{tcomb}}N_{\mathrm{rcomb}}^{K}\left[(KN_{\mathrm{ract}})^{3}+(KN_{\mathrm{ract}})^{2}\vphantom{+(KN_{r})^{2}\left(N_{\mathrm{tact}}+\frac{1}{2}\right)+(KN_{r})\left(N_{\mathrm{tact}}+\frac{3}{2}\right)+3)}\right.\\
 & \left.\vphantom{N_{\mathrm{tcomb}}((KN_{r})^{3}}\times\left(N_{\mathrm{tact}}+\frac{1}{2}\right)+(KN_{\mathrm{ract}})\left(N_{\mathrm{tact}}+\frac{3}{2}\right)+2\right].
\end{align*}

\subsection{Hardware Complexity}

The fact that the transmitted \ac{IQ} streams are received by all
the receive antennas, and not by some receive antenna subset, enables
\ac{DL-TR-GSM} to use a smaller number of receive antennas compared
to \ac{MU-TR-GSM}. As the number of receive antennas corresponds
to the number of RF chains at the receiver in RSM-based systems, the
hardware complexity advantage of \ac{DL-TR-GSM} becomes apparent.
Consequently, we can illustrate it in terms of the receive power consumption.
In the following, the receive power consumption is expressed relative
to the low noise amplifier power $P_{\mathrm{LNA}}$, the RF chain
power $P_{\mathrm{RFC}}$, the analog-to-digital converter power $P_{\mathrm{ADC}}$
and the baseband power $P_{\mathrm{BB}}$. The receive power consumption
for both schemes may be computed as
\[
P_{\mathrm{TOT}}=N_{r}(P_{\mathrm{LNA}}+P_{\mathrm{RFC}}+P_{\mathrm{ADC}})+P_{\mathrm{BB}}.
\]
The component powers are expressed relative to the reference power
$P_{\mathrm{ref}}$ as $P_{\mathrm{LNA}}=P_{\mathrm{ref}}$, $P_{\mathrm{RFC}}=2P_{\mathrm{ref}}$
and $P_{\mathrm{ADC}}=P_{\mathrm{BB}}=10P_{\mathrm{ref}}$ \cite{Mendez-Rial2016}.
For the reception of 2 IQ streams per user, under the same data rate,
\ac{DL-TR-GSM} requires $N_{r}=2$ receive antennas, and \ac{MU-TR-GSM}
requires $N_{r}=4$ receive antennas from which $N_{\mathrm{ract}}=2$
receive antennas are always active. If $P_{\mathrm{ref}}=20\,\mathrm{mW}$
\cite{Mendez-Rial2016}, the receive power consumption for \ac{DL-TR-GSM}
and \ac{MU-TR-GSM} are 720\,mW and 1240\,mW, respectively. Hence,
\ac{DL-TR-GSM} requires 520\,mW less power per user, corresponding
to a 41.9\,\% reduction compared to \ac{MU-TR-GSM}. These power
gains may provide remarkable power savings, especially for communication
systems supporting a large number of users.

\section{Separate Detector}

Due to limited hardware and software resources in user terminals,
a direct implementation of the \ac{ML} detector in \prettyref{eq:ML_det}
may be unsuitable for practical use. Motivated by this fact, we propose
a low-complexity separate detector, which executes a two-step detection
process.

In the first step, the separate detector determines the most likely
transmit \ac{IQ} symbol vector and power level matrix for each transmit
antenna combination. Assuming that the \mbox{$s_{1}$-th} transmit
antenna combination is activated, we perform the following signal
processing at the receiver utilizing $\mathbf{U}_{s_{1}}^{(k)^{\mathrm{H}}}$as
\[
\tilde{\mathbf{y}}_{s_{1}}^{(k)}=\mathbf{U}_{s_{1}}^{(k)^{\mathrm{H}}}\mathbf{y}^{(k)}=\mathbf{U}_{s_{1}}^{(k)^{\mathrm{H}}}\mathbf{G}_{s}^{(k)}\mathbf{P}_{i}^{(k)}\mathbf{\tilde{x}}_{m}^{(k)}+\mathbf{U}_{s_{1}}^{(k)^{\mathrm{H}}}\mathbf{n}^{(k)}.
\]
For $s=s_{1}$ (i.e. $\mathbf{U}_{s_{1}}^{(k)^{\mathrm{H}}}\mathbf{U}_{s}^{(k)}=\mathbf{I}_{N_{r}}$),
this system of equations corresponds to a set of $N_{r}$ parallel
subchannels. Hence, we can detect the power level and the transmission
symbol independently for each receive antenna $r\,(r=1,\dots,N_{r})$
as
\begin{multline}
\{\hat{m}_{s_{1}}(r),\hat{i}_{s_{1}}(r)\}=\\
\mathop{\mathop{\text{arg min}}}_{\substack{m_{s_{1}}(r)\in\{1,...,M\}\\
i_{s_{1}}(r)\in\{1,2\}
}
}\left|\tilde{y}_{s_{1}}^{(k)}(r)-\lambda_{s_{1},r}^{(k)}\beta_{s}\sqrt{P_{i_{s_{1}}(r)}^{(k)}}\widetilde{x}_{m_{s_{1}}(r)}^{(k)}\right|^{2}.\label{eq:sep_det-1-1}
\end{multline}
Now for each transmit antenna combination $s_{1}$ we have the candidate
power level matrix $\mathbf{P}_{i_{s_{1}}}^{(k)}=\mathrm{diag}(\sqrt{P_{i_{s_{1}}(1)}^{(k)}}\;\cdots\;\sqrt{P_{i_{s_{1}}(N_{r})}^{(k)}})$
and the candidate transmitted \ac{IQ} symbol vector $\mathbf{\tilde{x}}_{m_{s_{1}}}^{(k)}=\left[\widetilde{x}_{m_{s_{1}}(1)}^{(k)}\;\cdots\;\widetilde{x}_{m_{s_{1}}(N_{r})}^{(k)}\right]^{\mathrm{T}}$.

In the second step, we determine the active transmit antenna combination
$\hat{s}$ according to the following expression:
\begin{equation}
\hat{s}=\mathop{\mathop{\text{arg min}}}_{\substack{s\in\{1,...,N_{\mathrm{tcomb}}\}}
}\left\Vert \mathbf{y}^{(k)}-\mathbf{U}_{s}^{(k)}\mathbf{\mathbf{\Lambda}}_{s}^{(k)}\beta_{s}\mathbf{P}_{i_{s}}^{(k)}\mathbf{\tilde{x}}_{m_{s}}^{(k)}\right\Vert ^{2},\label{eq:sep_det-2-1}
\end{equation}
where $\mathbf{P}_{i_{s}}^{(k)}$ and $\mathbf{\tilde{x}}_{m_{s}}^{(k)}$
are obtained from the previous expression. Finally, we obtain the
transmitted bit sequence by combining the results from \prettyref{eq:sep_det-1-1}
and \prettyref{eq:sep_det-2-1}.

\section{Simulation Results}

In this section, we present the \ac{BER} simulation results of \ac{DL-TR-GSM}
with the \ac{ML} detector and the separate detector\@. As a benchmark,
we use \ac{MU-TR-GSM} for performance comparison. Then we show the
influence of the power ratio $\alpha$ on the \ac{BER} of \ac{DL-TR-GSM}.
Finally, we provide a comparison of the computational complexity of
\ac{DL-TR-GSM} and \ac{MU-TR-GSM}. 
\begin{figure}[t]
\centering{}\includegraphics{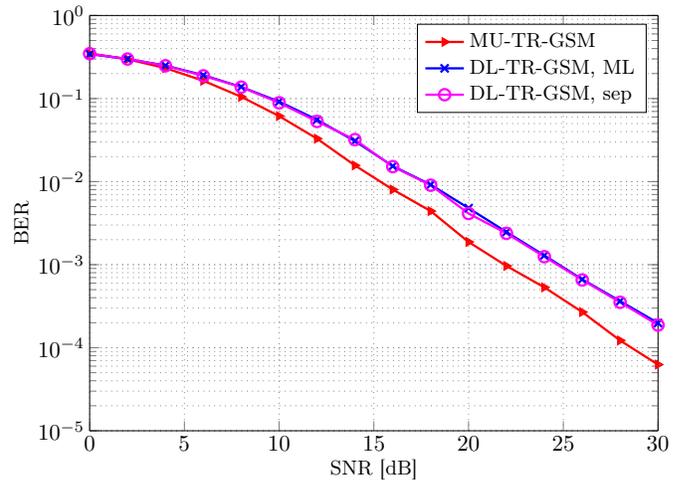}\caption{BER comparison without scaling coefficient $\beta_{s}$ (i.e., setting
$\beta_{s}=1$). \label{fig:BER-without}}
\end{figure}

We consider a downlink communication system with a single base station
and $K=2$ users. The base station has $N_{t}=32$ available transmit
antennas and always activates $N_{\mathrm{tact}}=4$ transmit antennas.
Each user receives 2 \ac{IQ} streams of QPSK symbols. To maintain
the same data rate, each user in \ac{DL-TR-GSM} is equipped with
$N_{r}=2$ receive antennas, while in MU-TR-GSM each user is equipped
with $N_{r}=4$ receive antennas from which $N_{\mathrm{ract}}=2$
receive antennas are always used for the IQ stream reception.
\begin{figure}[t]
\centering{}\includegraphics{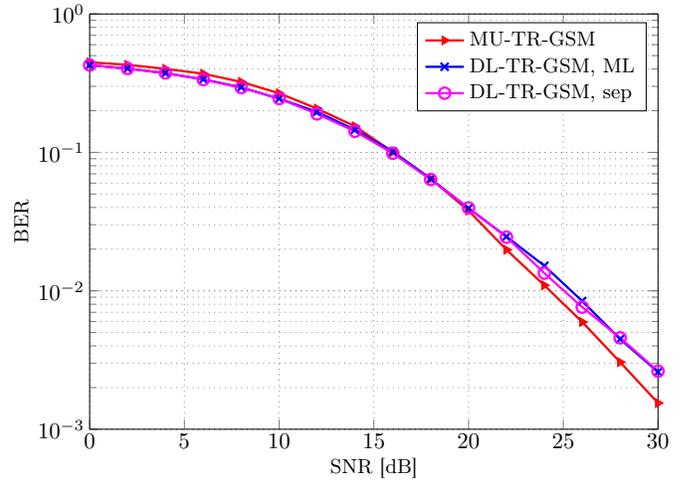}\caption{BER comparison with scaling coefficient $\beta_{s}$. \label{fig:BER-with}}
\end{figure}

In \prettyref{fig:BER-without}, we show the \ac{BER} of \ac{DL-TR-GSM}
and \ac{MU-TR-GSM} without utilizing the scaling coefficient $\beta_{s}$
(i.e., setting $\beta_{s}=1$). Due to omitting the scaling coefficient,
the precoder cannot maintain the same average signal power at the
input and the output. Hence this performance comparison cannot be
considered as generally fair, but we present it in order to show that
the BER performance matches with the results in \cite[Fig. 2]{Pizzio2016}.
In this case we define the \ac{SNR} as the ratio of the signal power
at the precoder \emph{input} and the noise power. This differs from
the standard way of defining the \ac{SNR} as the ratio of the signal
power at the transmit/receive antennas and the noise power.  In general,
\ac{DL-TR-GSM} achieves worse \ac{BER} than \ac{MU-TR-GSM} and
this effect becomes more pronounced at higher \ac{SNR}. Accordingly,
MU-TR-GSM exhibits up to 3\,dB lower \ac{BER} than \ac{DL-TR-GSM}
at high \ac{SNR}. On the other hand, both schemes have the same \ac{BER}
in the low-\ac{SNR} regime. As for the separate detector of \ac{DL-TR-GSM},
we see that it exhibits no performance loss with respect to the optimal
ML detector.

As already mentioned, omitting the scaling coefficient $\beta_{s}$
the precoders of the considered schemes are not able to maintain the
same average signal power and the \ac{BER} comparison in \prettyref{fig:BER-without}
cannot be classified as fair. Motivated by this, we present in \prettyref{fig:BER-with}
the \ac{BER} of \ac{DL-TR-GSM} and \ac{MU-TR-GSM} when the scaling
coefficient $\beta_{s}$ is utilized. In this case, the \ac{DL-TR-GSM}
shows a negligibly worse \ac{BER} performance than \ac{MU-TR-GSM}.\emph{
}Actually, the only visible difference is in the high-\ac{SNR} regime.
On the other hand, \ac{DL-TR-GSM} can even achieve slightly better
results at low \ac{SNR}. Again, the \ac{BER} of \ac{DL-TR-GSM}
remains extremely similar for the \ac{ML} detector and the separate
detector.

\begin{figure}[t]
\centering{}\includegraphics{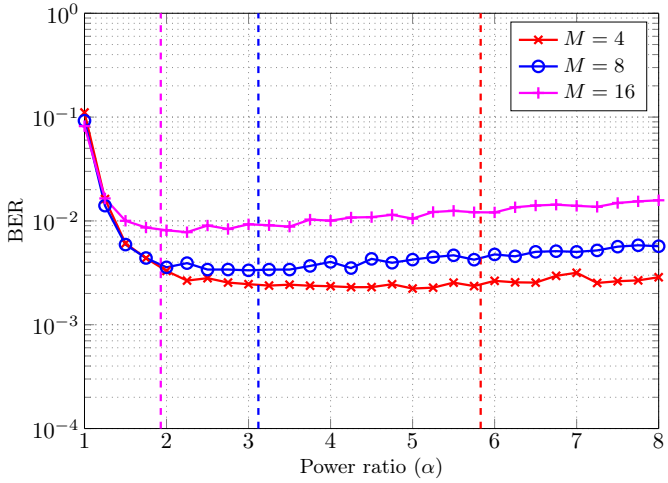}\caption{BER versus power ratio $\alpha$ for $\mathrm{SNR=30\,\mathrm{dB}}$.
Dashed lines present the optimum $\alpha$ obtained from \prettyref{eq:alpha_opt}.
\label{fig:BER-alpha}}
\end{figure}
To evaluate the correctness of the derived expression \prettyref{eq:alpha_opt},
we show the \ac{BER} of \ac{DL-TR-GSM} as a function of the power
ratio $\alpha$ in \prettyref{fig:BER-alpha}. The setup and parameters
are the same as in the previous figures, and the only difference is
that the IQ modulation order is allowed to vary. For the used IQ modulation
orders of 4, 8 and 16, the optimal values of $\alpha$ in \prettyref{eq:alpha_opt}
are 5.83, 3.12 and 1.93, respectively. These values are presented
with dashed lines in \prettyref{fig:BER-alpha}. In all cases the
optimal $\alpha$ in \prettyref{eq:alpha_opt} provides a BER which
is very close to the minimum achievable as shown by the simulations
in \prettyref{fig:BER-alpha}.  Also, it can be observed that the
\ac{BER} is very robust to the change of $\alpha$ for low IQ modulation
orders (e.g., $M=4$). 

\begin{figure}[t]
\centering{}\includegraphics{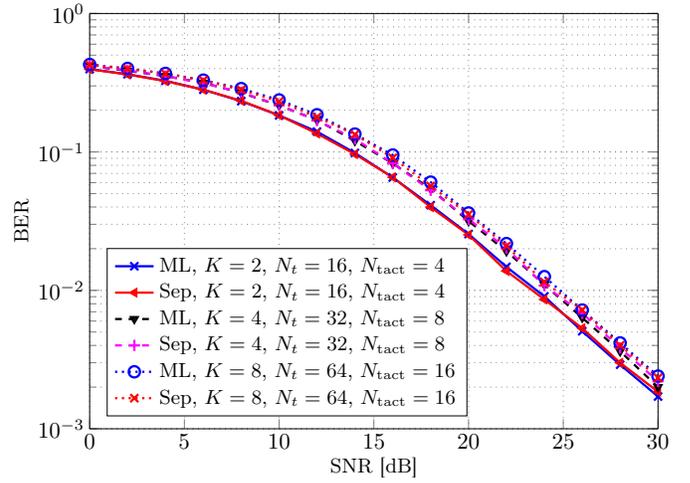}\caption{BER for different numbers of users. \label{fig:BER-var-K}}
\end{figure}
\begin{figure}[t]
\centering{}\includegraphics{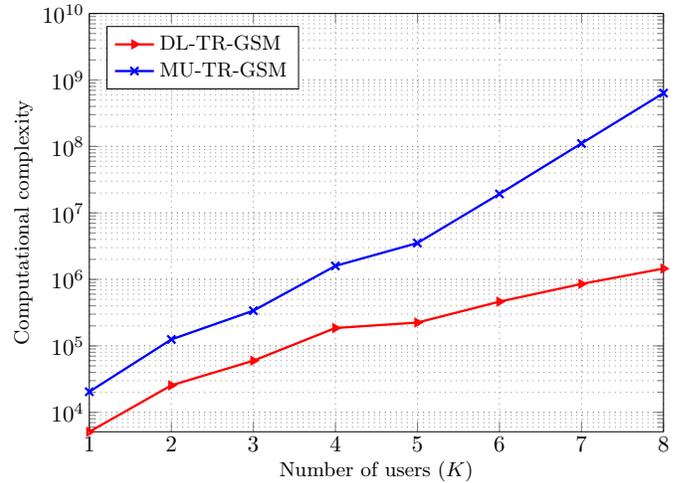}\caption{Computational complexity versus number of users $K$. \label{fig:Complexity}}
\end{figure}
In \prettyref{fig:BER-var-K}, we show the \ac{BER} of \ac{DL-TR-GSM}
for different numbers of users. Here we consider a number of users
$K$ equal to 2, 4 and 8, and we assume for each value of $K$ that
the number of active transmit antennas is $N_{\mathrm{tact}}=2K$
(assuming two receive antennas per user). Also, to maintain the same
data rate in the transmit spatial domain as $K$ increases, we assume
that the base station is equipped with $8K$ transmit antennas. It
can be seen that the \ac{BER} increases slightly with increasing
$K$, but this trend exhibits saturation at moderate values of $K$.
Therefore, in a system with many users, a minor variation in the number
of users has a negligible impact on the system's \ac{BER} performance.
Also, a good match is again observed between the \ac{BER} of \ac{DL-TR-GSM}
with the \ac{ML} detector and the \ac{BER} of \ac{DL-TR-GSM} with
the separate detector.

In \prettyref{fig:Complexity}, we present a computational complexity
comparison of \ac{DL-TR-GSM} and \ac{MU-TR-GSM} for a varying number
of users. We observe that \ac{DL-TR-GSM} is capable of achieving
a significant complexity reduction compared to \ac{MU-TR-GSM}\emph{.
}The main reason for this is the ability of \ac{DL-TR-GSM} to reduce
the number of \ac{SVD} computations and more notably the number of
$\beta_{s}$ values (note that the number of different scaling coefficient
values increases exponentially with $K$ for \ac{MU-TR-GSM}). Hence,
\ac{DL-TR-GSM} has the potential to provide a very significant computational
complexity reduction in real communication systems with a large number
of users.

\balance

\section{Conclusion}

In this paper, we proposed a new multiuser MIMO scheme, referred to
as \ac{DL-TR-GSM}, based on the concept of \ac{DLT}. In contrast
to \ac{MU-TR-GSM} which applies an activation of a subset of the
receive antennas, \ac{DL-TR-GSM} uses different power levels to transfer
information in the receive spatial domain. This operational change
provides a considerable complexity reduction in multiuser downlink
communications. For the same reason, the hardware complexity of the
user terminals decreases, causing a potentially large power saving.
To further improve the performance of \ac{DL-TR-GSM}, we proposed
a separate detector which reduces the detection complexity, while
maintaining a near-optimal BER.

\bibliographystyle{IEEEtran}
\phantomsection\addcontentsline{toc}{section}{\refname}\bibliography{IEEEabrv,IEEEexample,DL-TR-GSM}

\end{document}